\documentclass[12pt]{iopart}

\usepackage{graphicx}
\usepackage{layouts}
\usepackage[utf8]{inputenc}
\usepackage{graphicx}
\usepackage{array}
\usepackage{cite}
\usepackage{float}
\expandafter\let\csname equation*\endcsname\relax
\expandafter\let\csname endequation*\endcsname\relax
\usepackage{amsmath,amssymb}
\usepackage{iopams}
\usepackage{IEEEtrantools}
\usepackage{hyperref}
\usepackage{xcolor}
\usepackage{tikz}
\usetikzlibrary{quantikz2}
\usepackage{adjustbox}
\usepackage{arydshln}
\usepackage{mathtools}

\usepackage[normalem]{ulem} 

\usepackage[page]{appendix}  

\newcommand{\sx}{\sigma_x}

\graphicspath{ {./Figures/} }

\begin{document}


\title{Multi-Mode Global Driving of Trapped Ions for Quantum Circuit Synthesis}
\author{Philip Richerme$^{1,2,\dagger}$}
\address{$^1$Indiana University Department of Physics, Bloomington, Indiana 47405, USA}
\address{$^2$Indiana University Quantum Science and Engineering Center, Bloomington, Indiana 47405, USA}
\address{$^\dagger$richerme@iu.edu}
\begin{abstract}
We study the use of global drives with multiple frequency components to improve the efficiency of trapped ion quantum simulations and computations. We show that such `multi-mode' global drives, when combined with a linear number of single-qubit rotations, generate universal Ising-type interactions with shorter overall runtimes than corresponding two-qubit gate implementations. Further, we show how this framework may be extended to efficiently generate $n-$body interactions between any subset $n$ of the ion qubits. Finally, we apply these techniques to encode the Quantum Fourier Transform using quadratically-fewer entangling operations, with quadratically smaller runtime, compared with traditional approaches.

\end{abstract}
\maketitle

\section{Introduction}
Quantum computation and quantum simulation rely upon controlled physical processes to transform an initial quantum state into an output state. Such transformations are typically decomposed as discrete quantum gates \cite{nielsen2010quantum}, blocks of analog evolution under the Schr\"odinger equation \cite{monroe2021programmable}, or combinations of the two \cite{lanyon2011universal,parra2020digital}. Particularly in this era of noisy and intermediate-scale quantum (NISQ) devices \cite{preskill2018quantum}, finding decompositions that minimize the runtime of quantum algorithms, as well as the required number of entangling operations, is key to improving overall process fidelities and reducing the overhead required for quantum error correction \cite{steane2003overhead}.

The optimal decomposition of quantum algorithms depends sensitively on the capabilities and native instruction set of a chosen hardware platform. Although it is well-known that any algorithm may be decomposed into single-qubit and controlled-NOT gates \cite{divincenzo1995two,nielsen2010quantum}, alternative gate sets \cite{barenco1995elementary,lloyd1995almost,dodd2002universal,nielsen2010quantum} may be more natural to implement on specific quantum hardware. For instance, trapped-ion platforms often employ combinations of single-qubit rotations \cite{ballance2016high,gaebler2016high} and locally-addressed Ising-type entangling operations \cite{molmer1999multiparticle} to achieve universal quantum computation \cite{debnath2016demonstration,postler2022demonstration,moses2023race}. Moreover, global entangling operations may be realized with relative ease in trapped-ion systems \cite{lanyon2011universal,monroe2021programmable}, which may lead to improved efficiencies when implementing specific quantum algorithms or simulations \cite{martinez2016compiling,maslov2018use,hempel2018quantum,groenland2020signal,bravyi2022constant,schwerdt2022comparing,nemirovsky2025efficient}.

Global entangling operations have been used in most of the trapped ion Hamiltonian simulation experiments to date \cite{monroe2021programmable}. Typically, global laser beams are used to generate Ising-type spin-spin interactions that decay algebraically with distance \cite{kim2009entanglement}. However, much broader classes of interactions are achievable by applying multiple frequency components to the global beams, thereby engineering the couplings between the driving field and the ions' vibrational modes \cite{kyprianidis2024interaction,shapira2023programmable}. Although quantum simulations with such `multi-mode' global drives can directly address specific problems in quantum materials and quantum chemical systems \cite{kyprianidis2024interaction}, they cannot address arbitrary problems since they do not form a universal gate set on their own.

In this work we study the universal gate set comprised of multi-mode global drives and single-qubit rotations. We show how arbitrary Ising-type interactions may be generated by alternating layers of global drives with single-qubit rotations, with the number of layers scaling linearly with the number of qubits. In all studied cases, we find the runtime of this approach to be shorter than the corresponding implementation using direct two-qubit gates. Next we apply this technique to the generation of $n-$body interactions, again finding a relative speedup compared with the two-qubit gate approach. Finally we demonstrate how the Quantum Fourier Transform may be decomposed using multi-mode global drives and single-qubit rotations, providing a quadratic improvement in both the overall runtime and number of required entangling operations. 

The article is structured as follows. Section \ref{framework} reviews the background of generating spin-spin interactions using global laser beams and the types interactions which may be generated by controlling the couplings to the ion vibrational modes. We then introduce a template circuit based on multi-mode global drives that provides universal computation and simulation. In section \ref{ArbIsing}, we apply this template circuit to generate arbitrary Ising-type models and characterize its performance in implementing power-law interactions, spin-glass systems, and random matrices. Section \ref{MultiQubit} focuses on the implementation and performance of $n$-qubit gates, with $n > 2$, while section \ref{QFT} discusses the composition of the Quantum Fourier Transform using our gateset of multi-mode drives and single-qubit rotations. We conclude in section \ref{conclusion} with a discussion of our results, and possible challenges and opportunities for experimental implementation.


\section{Universal Quantum Operations Using Multi-Mode Global Drives}
\label{framework}

\subsection{Generating Effective Spin-Spin Interactions}
We consider a collection of $N$ trapped-ion qubits, with electronic basis states $\ket{\downarrow}_z$ and $\ket{\uparrow}_z$, confined in a global potential with a set of $N$ transverse mode frequencies $\omega_k$ ($k=1,\ldots,N$). Effective spin-spin interactions between ions may then be generated by driving with radiation at a frequency close to the mode frequencies $\omega_k$ \cite{wineland1998experimental,blatt2008entangled,monroe2021programmable}. When ions are addressed using a bichromatic electric field of the form $\Vec{E}=E_0\hat{y}\cos[kx-(\omega_0\pm \mu) t+\phi]$, the resulting laser-ion Hamiltonian may be written \cite{wineland1998experimental}:
\begin{equation}\label{eq:MotherEqnMultiIon}
H_\text{phys} = \sum_i^N-d_i E_0 \sigma_x^i \cos (kx_i-\omega_0 t \pm \mu t+\phi)
\end{equation}
where $d_i$ is the magnitude of the electric dipole operator for the $i$th ion, $\sx^i$ is the Pauli spin flip operator between the $\ket{\downarrow}_z$ and $\ket{\uparrow}_z$ states on ion $i$, and $\mu$ is the detuning of the exciting radiation from the qubit splitting $\omega_0$. 

In the regime where the motional modes are only virtually excited, or at times when the ions' phase space trajectories have all closed, the time evolution under $H_\text{phys}$ may be approximated by evolution under an effective Ising-type Hamiltonian \cite{kim2009entanglement}:
\begin{equation}\label{eq:SpinHamiltonian}
H_\text{Ising} = \sum_{i<j} J_{ij} \sx^i \sx^j
\end{equation}
where the strength of coupling between ions $i$ and $j$ is given by
\begin{equation}\label{eq:Jij}
J_{ij} = \Omega_i \Omega_j R\sum_k^N \frac{B_{ik}B_{jk}}{\mu^2-\omega_k^2}
\end{equation}
In Eq. \ref{eq:Jij}, $\Omega_i$ is the on-resonance Rabi frequency at ion $i$, $R$ is the recoil frequency $R=\hbar(\Delta k)^2/(2m)$, and $\hbar\Delta k$ is the momentum transfer from the electric field to each ion. The summation is performed over all mode indices $k$ and depends upon the normal mode matrix elements $B_{ik}$ of the mode vectors $\vec{b}_k$, as well as the corresponding mode frequencies $\omega_k$. When only two ions $i$ and $j$ have non-zero Rabi frequency, this formulation is equivalent to the arbitrary-angle M{\o}lmer-S{\o}rensen gate $MS(\chi_{ij})$ \cite{molmer1999multiparticle}, where the gate angle $\chi_{ij}=J_{ij}T$, and $T$ is the gate time. Conversely, when a global beam illuminates all ions simultaneously, this generates simultaneous fully-connected interactions between all pairs of ion qubits.

Equation \ref{eq:Jij} may be generalized to account for driving fields with $M$ bichromatic tones, each with independent detunings $\mu_m$ and global Rabi frequencies $\Omega_m$. Already, drives with multiple frequency components have been utilized to generate robust entangling gates \cite{shapira2018robust,manovitz2022trapped} and broaden the classes of quantum simulation experiments accessible to trapped-ion systems \cite{shapira2020theory,shapira2023programmable,kyprianidis2024interaction}. We refer to this condition as `multi-mode' global driving, generating the following coupling strength between ions $i$ and $j$:
\begin{equation}\label{eq:mmJij}
J_{ij} = \sum_k^N \sum_m^M \frac{\Omega_m^2 R}{\mu_m^2-\omega_k^2}B_{ik}B_{jk}
\end{equation}
Following \cite{kyprianidis2024interaction}, Eq. \ref{eq:mmJij} may be rewritten as
\begin{equation}\label{eq:LinearCombinationJks}
J_{ij} = \sum_k^N c_k J_{ij}^{(k)}.    
\end{equation}
where the set of matrices $J^{(k)}\equiv \vec{b}_k \otimes \vec{b}_k$ depends only on the mode vectors $\vec{b}_k$ and are weighted by a coefficient $c_k = \sum_m \Omega_m^2 R/(\mu_m^2-\omega_k^2)$. Since the $J^{(k)}$ matrices are dictated exclusively by the trapping potentials, the only tunable knobs for changing the interactions between ions are the mode frequencies $\mu_m$ and amplitudes $\Omega_m$, which comprise the weights $c_k$. Through numerical optimization, one may determine a set of $\Omega_m$ amplitudes which yield any desired $c_k$ weights, while satisfying all mode phase-space closure constraints and minimizing the evolution time $T$ (subject to the available laser power). We take this procedure as a prerequisite for the methods introduced below. 

\begin{figure*}[t]
\centering
\includegraphics[width=.8\textwidth]{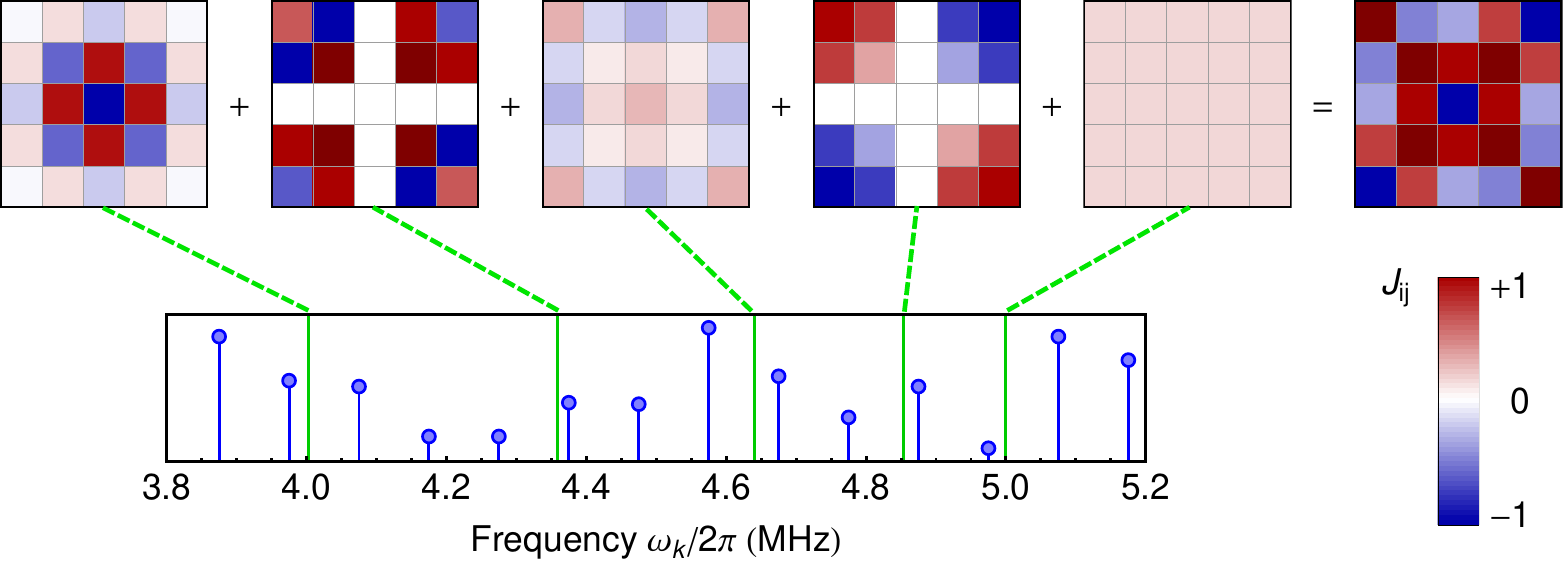}
\caption{
Transverse modes of $N=5$ harmonically-confined ions in a one-dimensional chain. Green lines show the mode frequencies. These modes may be driven using multiple bichromatic tones with independent amplitudes (blue lines). Insets show each normal mode matrix $J^{(k)}$, scaled by its contribution $c_k$, which add up to produce the final $J_{ij}$ couplings (rightmost matrix).
}
\label{fig:modespectrum}
\end{figure*}

This formalism is shown graphically in Fig. \ref{fig:modespectrum}, where multiple independent tones are used to drive a crystal of $N=5$ ions. The $N=5$ $J^{(k)}$ matrices, corresponding to each of the normal modes, inherit the structure of the normal mode vectors and are displayed above the spectrum. The matrices are each weighted by their coefficient $c_k$ and summed together to generate the final coupling $J_{ij}$ between all pairs of spins. As explored in \cite{kyprianidis2024interaction} and \cite{shapira2020theory}, this ability to arbitrarily tune the mode weights $c_k$ can lead to novel classes of trapped-ion quantum simulation experiments using exclusively global beams.

\subsection{Template Protocol for Universal Quantum Operations}
\label{sec:template}
Although multi-mode global drives can expand the types of interactions available to trapped-ion qubits, they are not universal for quantum computation or simulation. For example, Ising-type models (Eq. \ref{eq:SpinHamiltonian}) with arbitrary $J_{ij}$ couplings cannot be constructed using multi-mode global drives alone. This is because fully-connected Ising models contain $\mathcal{O}(N^2)$ pairwise interactions between $N$ qubits, but multi-mode global drives can at most specify $\mathcal{O}(N)$ couplings via the normal-mode weights $c_k$. More degrees of freedom are therefore required to achieve arbitrary interactions.

\begin{figure*}[b]
\centering
\begin{quantikz}[row sep=0.2cm,column sep=0.3cm]
& \gate[4]{U_0} &\gate{Z} & \gate[4]{U_1} & \gate{Z} &          & \gate[4]{U_2} &          &\\
&               &         &               &          & \gate{Z} &               & \gate{Z} &\\
&               &         &               &          &\ghost{Z} &               &\ghost{Z} &\\
&               &         &               &          &\ghost{Z} &               &\ghost{Z} &
\end{quantikz}
$\cdots$
\begin{quantikz}[row sep=0.2cm,column sep=0.3cm]
&\ghost{Z} & \gate[4]{U_N} &         &\\
&\ghost{Z} &               &         &\\
&\ghost{Z} &               &         &\\
&\gate{Z}  &               & \gate{Z}&
\end{quantikz}
\caption{
Template circuit showing the general framework used in this paper. Global drives enact the Unitary $U_n=e^{-iH_n t_n}$ for each analog block $n$, while pairs of $Z$ gates sequentially flip each of the $N$ qubits. 
}
\label{fig:templatecircuit}
\end{figure*}
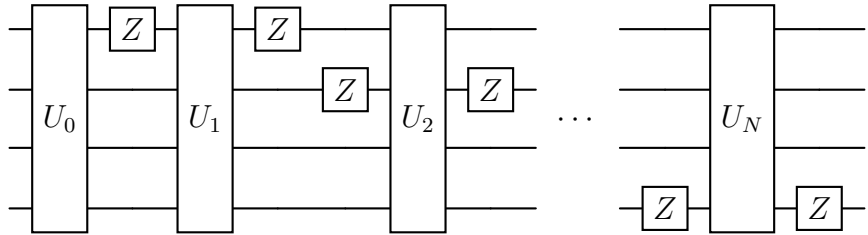

In this work, we introduce the template circuit in Fig. \ref{fig:templatecircuit} as a means of achieving universal quantum operations. This circuit, which is comprised of multi-mode global drives and single-qubit rotations, combines the techniques of hybrid digital-analog quantum simulations \cite{arrazola2016digital,parra2020digital} with the multi-mode global drives first explored in \cite{kyprianidis2024interaction}. In this protocol, $N+1$ blocks of analog time evolution are interspersed with pairs of single-qubit $R_z(\pi)$ rotations, labeled $Z$ in Fig. \ref{fig:templatecircuit}. Within each analog block, unitary evolution $U_n=e^{-i H_n t_n}$ proceeds under the Hamiltonian $H_n = \sum_{i<j} J_{ij}^{n} \sigma_i^x\sigma_j^x$ for time $t_n$, where the couplings $J_{ij}^{n} = \sum_k c_k^n J_{ij}^{(k)}$ in each block are adjusted by controlling the mode weights $c_k^n$. Further, the pair of $R_z(\pi)$ rotations surrounding each block flip the spin of a single ion, inverting its sign of coupling in the overall $J_{ij}^{n}$ matrix. Such spin flips allow for the generation of $J_{ij}^{n}$ matrices outside the accessible domain of global beams, and they may be viewed analogously to spin-echo pulses for canceling unwanted couplings between qubits. 

Compared to traditional digital-analog approaches, multi-mode global drives provide a quadratic reduction in the number of analog blocks and single qubit flips required for arbitrary interactions. In \cite{parra2020digital}, for instance, each analog evolution block evolves a fixed Hamiltonian $H=\sum_{i<j} J_{ij} \sigma_i^x\sigma_j^x$ for a variable time $t_n$. Since the Hamiltonian is constant, arbitrary control over all $N(N-1)/2$ couplings requires $N(N-1)/2$ analog evolutions sandwiched by two-qubit pairs of $R_z(\pi)$ rotations. In contrast, multi-mode driving removes the restriction of fixed Ising couplings during each analog evolution block. As a result, this allows for the generation of $\mathcal{O}(N)$ linearly-independent Hamiltonians $H_n$; thus, only $\mathcal{O}(N)$ blocks sandwiched by single-qubit flips are required to achieve arbitrary interactions.

In the sections below, we will apply this multi-mode global drive framework to build complex Hamiltonians and quantum circuits of interest to quantum simulation and computation. The available control parameters will be the $N-1$ linearly-independent mode weights $c_k^n$ within each analog block\footnote{Since the $J^{(k)}$ matrices satisfy the property that $\sum_k J^{(k)}=\mathbb{I}$ \cite{kyprianidis2024interaction}, the vibrational modes provide only $N-1$ independent degrees of freedom.}, as well as the $N+1$ block evolution times $t_n$ in the template circuit. All together, these $(N-1)(N+1)$ parameters form an overcomplete set for generating arbitrary Ising interactions. This enables the freedom to select an optimal set of parameters $c_k^n$ and $t_n$ which minimize the overall circuit runtime, as will be explored in Sec. \ref{sec:arbising}.


\section{Arbitrary Ising Interactions}
\label{ArbIsing}

In this section, we will show how to generate Ising interactions (Eq. \ref{eq:SpinHamiltonian}) with arbitrary $J_{ij}$ couplings between spins, using the template circuit introduced in Sec. \ref{framework}. Full control over these couplings opens the possibility for trapped-ion quantum simulators to study, for instance, interacting spin models with \emph{exact} (rather than approximate) power-law interactions \cite{monroe2021programmable}, Ising spin-glass systems \cite{binder1986spin}, optimization problems \cite{farhi2014quantum,pagano2020quantum}, and effective higher-dimensional spin lattices \cite{rajabi2019dynamical,wu2023qubits}. In addition, arbitrary control over the Ising-type spin-spin couplings may be extended to quantum simulations of arbitrary XY or Heisenberg-type models, either through analog \cite{richerme2014non,bermudez2017long} or digital approaches \cite{lanyon2011universal}.

\subsection{Comparison to prior work}
Several existing approaches have been proposed and demonstrated for constructing arbitrary Ising-type couplings between trapped ions. In the `direct' approach, which will serve as a basis for comparison in the rest of this paper, two-qubit M{\o}lmer-S{\o}rensen gates are used to sequentially apply the couplings $\{J_{12},J_{13},\ldots,J_{N-1,N}\}$. This technique requires experimental hardware that can perform locally-addressed entangling operations, with $N(N-1)/2$ gates required to build the full $J_{ij}$ matrix. In principle, the implementation runtime of this technique is the two-qubit gate time multiplied by the number of non-zero $J_{ij}$ matrix elements. It has also been proposed to modify the direct approach by adding multiple frequency components to each laser beam, creating arbitrary interactions in parallel \cite{Korenblit2012}, or by implementing phase modulation to create a desired Ising interaction pattern \cite{lu2019global,lu2025implementing}. However, these approaches still require locally-addressed beams to generate entanglement.

To avoid the experimental challenges of locally-addressed entangling gates, several proposals have discussed ways to achieve complex couplings by combining global drives with locally-addressed single-qubit rotations. In \cite{hayes2014programmable}, for instance, local $R_z(\pi)$ pulses are used to transform an initial long-range Ising Hamiltonian with translation invariance into any other Hamiltonian in that class. Likewise, adding multiple frequency tones to global entangling beams has been shown to create a wide variety of interesting coupling patterns relevant to quantum many-body spin systems \cite{shapira2020theory,shapira2023fast}. Fully-arbitrary Ising couplings are made possible following the digital-analog methods in \cite{parra2020digital}, which intersperse $N(N-1)/2$ blocks of analog evolution under a fully-connected Ising Hamiltonian with double pairs of local $\pi$ spin flips. For this case, the overall runtime and number of entangling blocks both scale as $\mathcal{O}(N^2)$. It has also been shown that the implementation runtime can be reduced to linear $\mathcal{O}(N)$ scaling by allowing arbitrary numbers of qubits to be flipped at each digital layer \cite{bassler2023synthesis}. To determine the appropriate flips and sequence timing, a linear program was used to generate the target Hamiltonian in the shortest runtime, given an input Hamiltonian and variable analog evolution times between digital steps.

\subsection{Synthesizing Arbitrary Ising Interaction with Multi-Mode Global Drives}
\label{sec:arbising}
Here, we outline our approach for generating arbitrary Ising interactions using multi-mode global drives and a linear number of single-qubit spin flips. Compared to prior efforts, we report three primary advantages of our method: (1) The template circuit in Fig. \ref{fig:templatecircuit} requires only $N$ single-qubit flips, rather than $N^2$ \cite{parra2020digital,arrazola2016digital} or more \cite{bassler2023synthesis}, which simplifies the overall circuit and reduces the overall single-qubit gate infidelity; (2) Finding the optimal solution in our approach only requires polynomial resources, rather than the exponential resources required in Ref. \cite{bassler2023synthesis}. As a result, we can easily find solutions for 100+ ions, well beyond prior demonstrations; (3) We observe faster quantum circuit runtimes for our method compared to direct two-qubit gate approaches as well as the approaches in \cite{parra2020digital,arrazola2016digital,bassler2023synthesis}, while requiring fewer single-qubit gates and classical optimization time. 

 To begin, we assume an ion trap quantum simulator operating in the Lamb-Dicke regime, capable of implementing the fully-connected Ising couplings as in Eq. \ref{eq:mmJij} as well as local $R_z(\pi)$ rotations. At the $n^{\text{th}}$ timestep in the template circuit (Fig. \ref{fig:templatecircuit}), pairs of $R_z(\pi)$ rotations flip the spin of the $n^{\text{th}}$ qubit during that block's analog evolution. This transformation is equivalent to evolution under the Hamiltonian $H_n' = \sum_{i<j} J_{ij}^{n'} \sigma_i^x\sigma_j^x$, where $J_{ij}^{n'}$ is the original $J_{ij}^{n}$ matrix with inverted coupling sign in the $n^{\text{th}}$ row and column. The full unitary evolution enacted by the template circuit in Fig. \ref{fig:templatecircuit} is then:
\begin{equation}
\label{eq:fullunitary}
    U=\prod_n \exp\left[-i H'_n t_n \right]=\exp\left[-i\sum_n H'_n t_n\right]=\exp\left[-i \sum_{i<j} \left(\sum_n J_{ij}^{n'} t_n \right) \sigma_i^x\sigma_j^x\right]
\end{equation}
Equation \ref{eq:fullunitary} demonstrates that the template circuit can be made to replicate the dyanmics of any desired 
Ising coupling $J_{ij}^\text{des}$, so long as $\left(\sum_n J_{ij}^{n'} t_n \right)$ can be made equivalent to $J_{ij}^\text{des}$.

Our goal will be to find the parameters which exactly replicate the target Ising couplings, while minimizing the overall runtime $\sum_n t_n$. It is critical that constraints be placed on the strength of the mode weights $c_k^n$; if not, the circuit runtimes can be made arbitrarily small by making the driving Rabi frequencies (and thereby the $c_k^n$) arbitrarily large. We therefore assume a finite laser power to be split between all $k$ normal modes, such that $\sum_k |c_k^n| = 1$ at each timestep $n$. Using Eq. \ref{eq:LinearCombinationJks}, the unitary evolution within each analog block may be written:

\begin{equation}
\label{eq:un}
    U_n = \exp\left[-i H_n t_n\right]=\exp\left[-i \sum_k \sum_{i<j} t_n c_k^n J_{ij}^{(k)}\sigma_i^x\sigma_j^x\right] = \exp\left[-i \sum_k \sum_{i<j} \tilde{c}_k^n J_{ij}^{(k)}\sigma_i^x\sigma_j^x\right]
\end{equation}
where in the last equality we have absorbed the evolution time into the weight constant, such that $\tilde{c}_k^n = t_n c_k^n$. Then, under the constraint $\sum_k |c_k^n| = 1$, minimizing the overall runtime $\sum_n t_n$ is equivalent to minimizing $\sum_n \sum_k |\tilde{c}_k^n|$, since 
\begin{equation}
\label{eq:cknruntime}
 \sum_n \sum_k |\tilde{c}_k^n| = \sum_n \sum_k t_n |c_k^n| = \sum_n t_n
\end{equation}

To find the optimal mode weights $\tilde{c}_k^n$, we first `vectorize' each $J_{ij}^{(k),n}$ matrix by arranging its $N(N-1)/2$ upper-triangular entries into a column vector, $\vec{j}_k^n$ \cite{parra2020digital}. Next, we build a matrix whose columns are the $\vec{j}_k^n$ vectors. At each timestep, there are $N-1$ linearly-independent $\vec{j}_k$ vectors, and there are $N+1$ timesteps, for a total of $(N-1)(N+1)$ columns. We seek that this matrix, times the column vector of all $\tilde{c}_k^n$ values, generates the vectorized version of the desired couplings, $\vec{j}_\text{des}$.

Since the number of $\tilde{c}_k^n$ elements form an overcomplete set compared to the $N(N-1)/2$ Ising couplings, we use a linear program \cite{karloff2008linear,bassler2023synthesis} to find the specific $\{\tilde{c}_k^n\}$ which minimize the circuit runtime, subject to the constraint of generating the desired couplings. Since the $\tilde{c}_k^n$ may themselves be positive or negative, care must be taken to ensure the minimization does not diverge. We address this issue by doubling the width of the matrix described above, explicitly including positive and negative versions of each column vector $\vec{j}_k^n$, as shown in the matrix structure below:

\begin{center}
$\left(\begin{array}{cccc:cccc}
	\vdots & \vdots & \vdots & \vdots & \vdots & \vdots & \vdots & \vdots\\ 
	\vec{j}_1^0 & \vec{j}_2^0 & \cdots & \vec{j}_{N-1}^N & -\vec{j}_1^0 & -\vec{j}_2^0 & \cdots & -\vec{j}_{N-1}^N \\
    \vdots & \vdots & \vdots & \vdots & \vdots & \vdots & \vdots & \vdots
\end{array}\right)\left(\begin{array}{c} \vdots \\ \tilde{c}_k^{n,+} \\ \vdots \\ \hdashline  \vdots \\ \tilde{c}_k^{n,-} \\ \vdots \end{array}\right)=\left(\begin{array}{c} \vdots \\ \vec{j}_\text{des} \\ \vdots\end{array}\right)$
\end{center}
This framework forces all $\tilde{c}_k^n \geq 0$, allowing a standard linear program to minimize the sum over all $2(N-1)(N+1)$ parameters $\tilde{c}_k^{n,\pm}$ while satisfying the matrix equation shown above. 

We implement the linear program using the \texttt{LinearProgramming[]} function in Mathematica. Inputs to this function are the $\vec{j}_k^n$ vectors, which are pre-determined based on the normal modes and single-qubit spin flips, and the desired coupling vector $\vec{j}_\text{des}$. This function outputs a list of mode weights for each mode $k$ at timestep $n$, where the position of each mode weight within the double-length column vector indicates whether a specific $\tilde{c}_k^n$ should be set positive or negative. Then, following Eq. \ref{eq:cknruntime}, the overall circuit runtime is calculated by summing $\sum_n \sum_k |\tilde{c}_k^n|$. 

Although the simplex method of linear programming might exhibit exponential worst-case complexity \cite{klee1972}, polynomial-time scaling with system size is more commonly observed in practice \cite{spielman2004smoothed}. For our application, the number of variables $\tilde{c}_k^n$ scales as $\mathcal{O}(N^2)$, suggesting that the linear programming approach requires an overall execution time polynomial in the number of qubits $N$. Indeed, the numerical results presented in the rest of this work (with up to $N = 100$ qubits) all took $< 1 $ minute per point to obtain, using a desktop computer with an Intel Core i7 Processor (3.6 GHz) and 64 GB of RAM.

\subsection{Numerical Results}
We now compare two approaches for generating arbitrary Ising interactions. In the first approach, local two-qubit entangling gates are applied sequentially to build the desired coupling matrix $J_{ij}^{des}$. We assume that these gates may be driven via the center-of-mass vibrational mode with perfect fidelity in a time $\tau$, and that couplings to all spectator modes are negligible. For a system with $k$ non-zero two-qubit couplings, the total implementation time is then $k \tau$ \cite{schindler2013quantum,figgatt2019parallel,shaffer2023sample}. In the second approach, global multi-mode drives are combined with single-qubit rotations (Fig. \ref{fig:templatecircuit}) to engineer the desired couplings. The overall implementation time is the sum of all analog-block runtimes, $\sum_n t_n$, which is equivalent to the sum of all effective mode weights $\sum_{n,k} |\tilde{c}_k^n|$ (as derived in Sec. \ref{sec:arbising}). We assume that single-qubit rotations can be performed much more quickly than entangling operations and contribute negligibly to the overall runtime. For all numerical results below, we define the \emph{scaled} runtime to be the total implementation time required to evolve the unitary $U=e^{-iHt}$ for (unitless) time $t=1$, given a desired Hamiltonian.

\begin{figure*}[t]
\centering
\includegraphics[width=.6\textwidth]{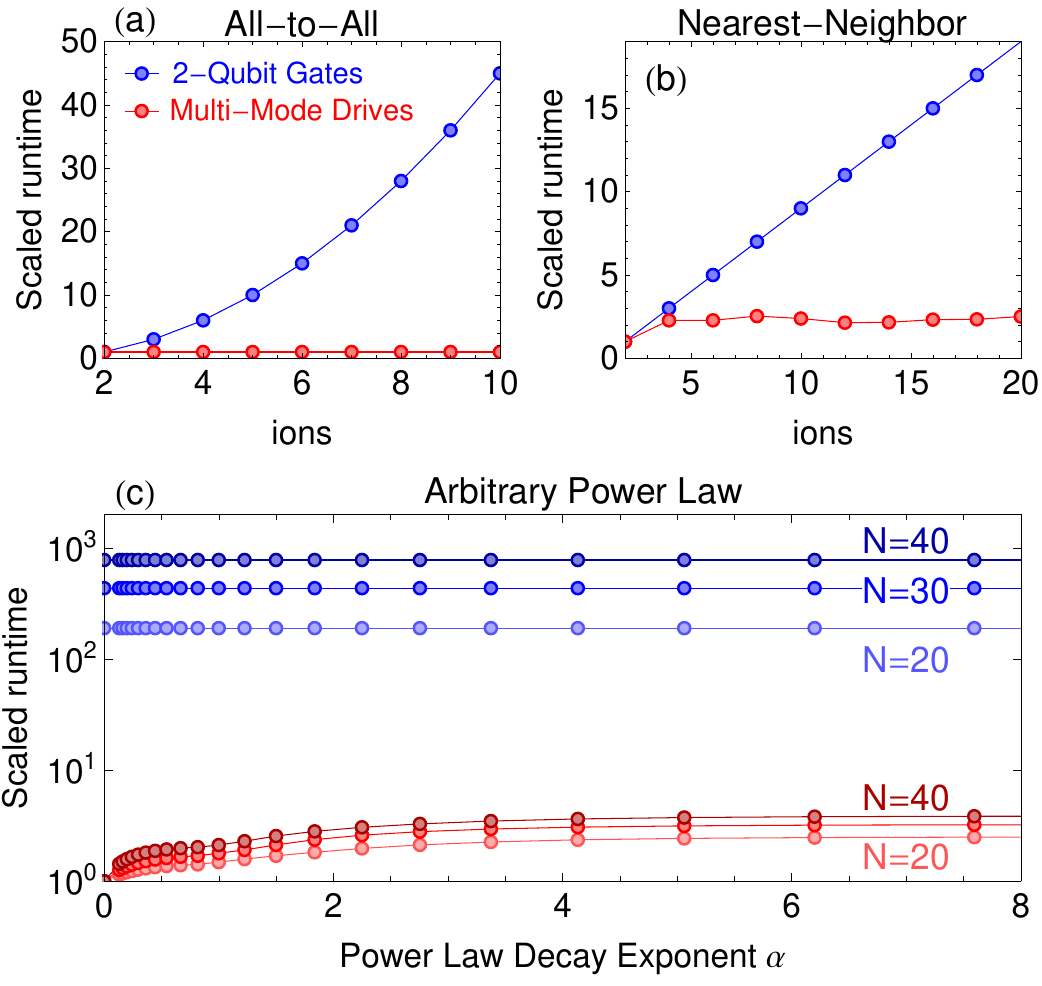}
\caption{
(a) The runtime for implementing all-to-all interactions using sequential two-qubit gates (blue) scales quadratically with ion number, compared to the constant-time scaling of multi-mode drives (red). (b) Nearest-neighbor-only interactions may also be implemented in shorter runtime using multi-mode drives. (c) Runtime comparison for Ising interactions with exact power-law decay, $J_{ij}=J_0/|i-j|^\alpha$, for $N=\{20, 30,40\}$ ions. In all cases, multi-mode drives (red) require significantly shorter implementation times than direct two-qubit gates (blue).
}
\label{fig:PowerLaw}
\end{figure*}

We start by investigating Ising couplings with long-range power-law interactions. To date, many quantum simulation experiments have generated approximate power-law couplings of the form $J_{ij}=J_0/|i-j|^\alpha$, with $0.5 < \alpha < 2$, using a single pair of bichromatic tones \cite{monroe2021programmable,islam2013emergence,richerme2014non}. Here, we explore models with exact power-law decay, over the full range $0 < \alpha < \infty$ (e.g. from all-to-all interactions to nearest-neighbor-only interactions).

In Fig. \ref{fig:PowerLaw}a, we compare the scaled runtimes for generating all-to-all couplings ($\alpha = 0$) using the direct two-qubit approach and the multi-mode drive approach. In the direct approach, there are $N(N-1)/2$ Ising couplings which must be individually implemented with sequential gates, leading to a quadratically-scaling runtime with system size. In contrast, multi-mode driving enables generation of all-to-all couplings by exciting the center-of-mass mode for a single timestep. The scaled runtime is therefore constant with system size. This multi-mode drive advantage becomes pronounced at even moderate system sizes, with a predicted 800x speedup for $N=40$.

For nearest-neighbor interactions ($\alpha = \infty$), multi-mode global drives show an approximately linear speedup compared to two-qubit gates. As shown in Fig. \ref{fig:PowerLaw}b, the direct approach requires $(N-1)$ Ising couplings to be implemented for the nearest-neighbor case, resulting in a linear scaling of runtime with system size. Simulations of the multi-mode drive approach exhibit approximately constant scaling. Figure \ref{fig:PowerLaw}c shows the runtimes for all power-law decays between all-to-all and nearest neighbor, for $N=\{20,30,40\}$ ions. For all ion numbers, and for all power law decay exponents, the multi-mode drives may be implemented with significantly shorter runtimes when compared to the sequential two-qubit gate approach.


\begin{figure*}[t]
\centering
\includegraphics[width=\textwidth]{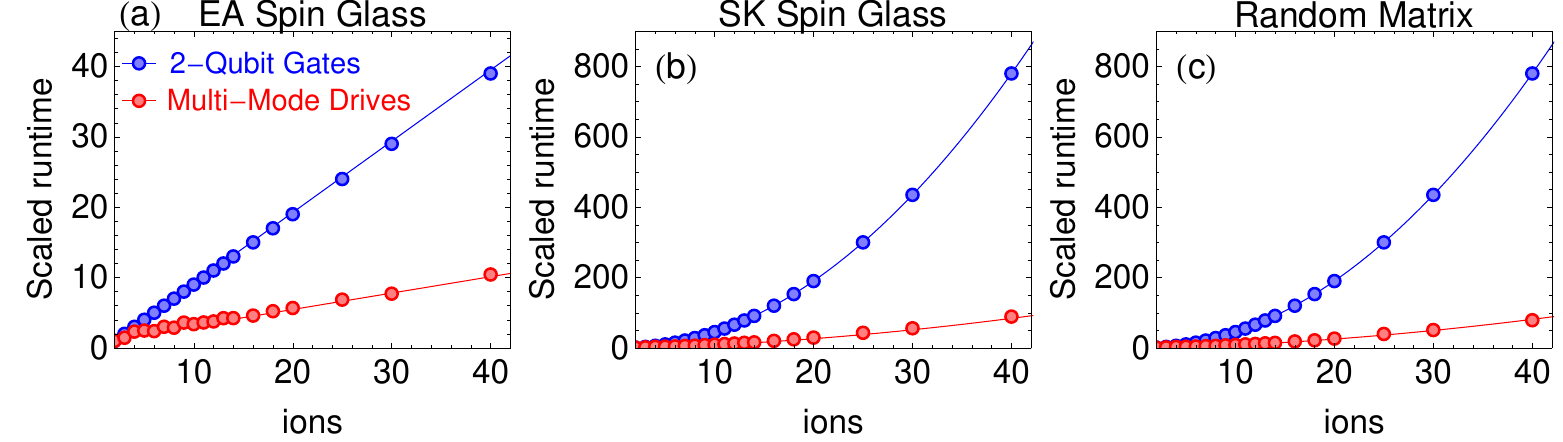}
\caption{
Scaled runtimes for sequential two-qubit gates (blue) and multi-mode drives (red) when averaging over random instances of (a) an Edwards-Anderson Ising spin glass, (b) a Sherrington–Kirkpatrick spin glass, and (c) a fully-randomized $J_{ij}$ interaction matrix.
}
\label{fig:RandomMx}
\end{figure*}

Similar runtime comparisons may be performed for Ising models with randomized $J_{ij}$ couplings. Such models may be used to characterize, for instance, spin glass magnetic systems in quantum condensed-matter physics \cite{binder1986spin}. In Fig. \ref{fig:RandomMx}a, we analyze the runtimes for implementing an Edwards-Anderson (EA) Ising spin glass \cite{edwards1975theory}, defined by nearest-neighbor-only interactions with Gaussian-distributed $J_{ij}$ of mean 0 and variance 1. Since there are $\mathcal{O}(N)$ random couplings in the EA model, the runtimes for both the direct two-qubit implementation and multi-mode drives exhibit linear scaling with system size, when averaged over 50 random instances per point. In all cases, the multi-mode drives outperform the direct implementation, due to the flexibility of choosing the mode weights $\tilde{c}_k^n$ from an overcomplete set.

We also analyze the performance of related models: a Sherrington–Kirkpatrick (SK) spin glass \cite{sherrington1975solvable}, which expands the EA model to include connections between all lattice sites, and a fully-random matrix model with uniformly distributed $J_{ij} \in [-1,1]$. As in Fig. \ref{fig:RandomMx}a, we average over 50 random iterations of each model. Figures \ref{fig:RandomMx}b-c demonstrate a quadratic scaling of runtime with ion number in both cases, resulting from the $\mathcal{O}(N^2)$ fully-connected couplings. Once more, we observe that the multi-mode drive protocol enables significantly shorter runtimes compared with the direct two-qubit gate approach. In addition, Fig. \ref{fig:RandomMx}c confirms that fully arbitrary Ising models are accessible using the simple template circuit introduced in Sec. \ref{sec:template}.

\subsection{Expanded Multi-Mode Drive Protocol}

\begin{figure*}[t]
\centering
\includegraphics[width=.8\textwidth]{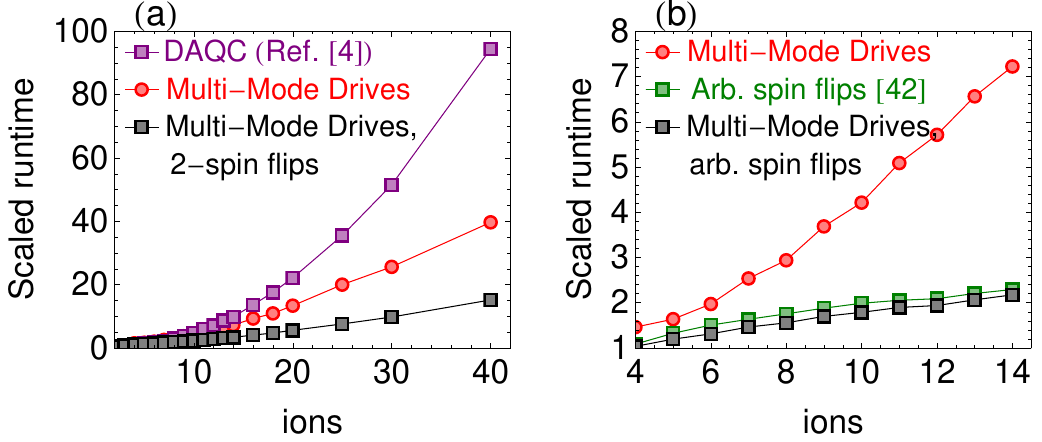}
\caption{
(a) The runtimes for simulating random $J_{ij}$ matrices by driving all-to-all interactions using pairs of single-qubit flips, as in Ref. \cite{parra2020digital}, are shown as purple squares. Adding multi-mode driving (black squares) improves the runtimes. Multi-mode driving using only sequential single-qubit flips (red circles), as in Fig. \ref{fig:RandomMx}(c), is shown for reference. (b) When arbitrary numbers of spin flips are allowed to bracket each Unitary block, as in \cite{bassler2023synthesis}, the predicted runtimes for simulating random interactions show further improvement. Green squares show driving with all-to-all interactions, and black squares show small improvement using multi-mode drives. Red circles again show multi-mode driving using only sequential single-qubit flips.
}
\label{fig:Comparisons}
\end{figure*}

Our multi-mode global drive protocol, as shown in Fig. \ref{fig:templatecircuit}, alternates $\mathcal{O}(N)$ analog evolution blocks with pairs of single-qubit spin flips. In contrast, protocols without multi-mode drives require $\mathcal{O}(N^2)$ blocks, surrounded by spin flips on 2 \cite{parra2020digital} to $N$ \cite{bassler2023synthesis} qubits in parallel. Here, we explore the reductions in runtime that are accessible by expanding our template circuit to include parallel $Z$ flips on multiple qubits between analog blocks.

In Fig. \ref{fig:Comparisons}a, we study the implementation runtimes for a fully-connected random Ising model, averaged over 50 random iterations. This model was discussed in Fig. \ref{fig:RandomMx}c and is shown again here (in red circles) for reference. Already, the multi-mode drive approach (with single $Z$ flips) outperforms single-mode drives with 2 parallel $Z$ flips \cite{parra2020digital}, shown as purple squares. Expanding the multi-mode drives to include 2 parallel $Z$ flips per block further improves the runtimes, shown as the black squares in Fig. \ref{fig:Comparisons}.

In Fig. \ref{fig:Comparisons}b, we study the same model while allowing for arbitrary numbers of parallel spin flips per block, from $1$ to $N$. In this case, single-mode drives with arbitrary numbers of flips (green squares) \cite{bassler2023synthesis} show improvement over the single-flip multi-mode drives (red circles). Expanding multi-mode drives to include arbitrary flips (black squares) shows only minor further improvement. Unfortunately, finding the optimal set of arbitrary spin flips scales exponentially in the system size, since there are are $2^N$ possible transformations of each $J_{ij}^{(k)}$ matrix per timestep. This approach therefore becomes infeasible for even modestly-sized lattices of a few dozen ions.


\section{$n$-Body Interactions}
\label{MultiQubit}
Multi-qubit gates, and $n$-body interactions, often arise naturally in the expression of quantum simulation and computation problems. For instance, many interesting systems in condensed matter physics, nuclear physics, and quantum chemistry contain $n$-body interactions as crucial components of the system the Hamiltonian \cite{whitfield2011simulation,casanova2012quantum,hauke2013quantum,yung2014transistor,du2023multinucleon}. In addition, it has been found that $n$-body interactions may be used to improve the efficiency of common quantum circuits compared to single- and two-qubit gates alone \cite{maslov2018use,nemirovsky2025efficient}. However, such interactions, which take the form $\sigma_\alpha^1 \otimes \sigma_\alpha^2 \otimes \ldots \otimes \sigma_\alpha^n$, $\alpha \in \{x,y,z\}$, require dedicated hardware implementations and in general cannot be synthesized using Trotterized sequences of two-body interactions \cite{lanyon2011universal}.  In this section, we will show how $n$-body interactions between trapped ions may be generated using the framework of multi-mode global driving, and we will demonstrate that the scaled runtime of this approach compares favorably to more traditional techniques. 

\subsection{Comparison to prior work}
To date, two primary approaches have been used to realize $n$-body interactions in trapped ion systems. In the first approach, $n$-body interactions are generated via spin-dependent squeezing, which arises from driving ions near their second set of motional sidebands \cite{katz2022n, katz2023programmable}. However, experimental implementations of this approach \cite{katz2023demonstration,zhukas2024observation} have required trapped-ion hardware with individually-addressed entanglement beams, which are outside the realm of global multi-mode drives.

\begin{figure*}[ht]
\centering
\begin{quantikz}[row sep=0.2cm,column sep=0.3cm]
& \gate[4]{e^{-iJ \sigma_1^x\sigma_2^x\sigma_3^x\sigma_4^x t}} &\\
& \ghost{Z}               &\\
& \ghost{Z}              &\\
& \ghost{Z}               &
\end{quantikz}
=
\begin{quantikz}[row sep=0.2cm,column sep=0.3cm]
&\gate{R(\frac{\pi}{2})} & \gate[4]{U_{MQ}(\frac{\pi}{2})} & \gate{R_z(\pm 2Jt)} & \gate[4]{U_{MQ}(-\frac{\pi}{2})} & \gate{R(-\frac{\pi}{2})} &\\
&\ghost{Z}               &               &              &                       & &\\
&\ghost{Z}               &               &              &                       & &\\
&\ghost{Z}               &               &              &                       & &
\end{quantikz}
\caption{
Template circuit for simulating the time evolution of a Hamiltonian with four-body interaction term $\sigma_1^x\sigma_2^x\sigma_3^x\sigma_4^x$. Rotation operations $R$ and global multi-qubit unitaries $U_{MQ}$ are defined in the text.
}
\label{fig:multiqubit}
\end{figure*}
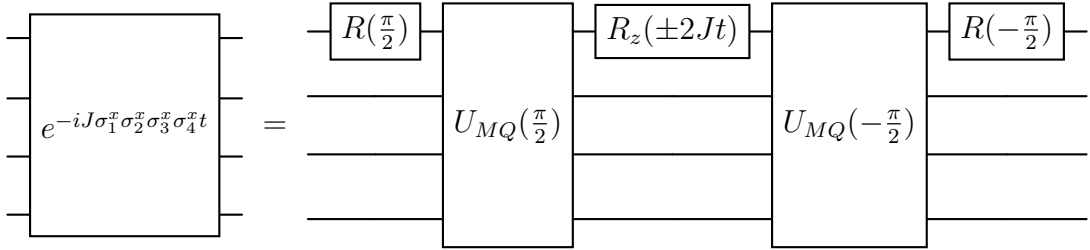

The second approach, described in Refs. \cite{lanyon2011universal} and \cite{Muller2011}, combines the all-to-all connectivity of global M\o lmer-S\o rensen operations with single-qubit rotations to create $n$-body gates. Following \cite{lanyon2011universal,Muller2011}, we show an example circuit for generating unitary time evolution under the 4-body Hamiltonian $H=J\sigma_x^1\sigma_x^2\sigma_x^3\sigma_x^4$ in Fig. \ref{fig:multiqubit}. Central to such circuits are the multi-qubit unitaries $U_{MQ}(\pm\pi/2)$, which are typically chosen to be fully-entangling M\o lmer-S\o rensen interactions with uniform couplings between all spins. These global interactions bracket a local $R_z$ rotation on the first qubit, with a rotation angle proportional to $2Jt$. (The positive rotation angle is used for $N=4k+2$ and $N=4k+3$ qubits, the negative angle is for $N=4k$ and $N=4k+1$ qubits, with $k \in \mathbb{N}$). Finally, the first qubit is subjected to pre- and post-rotations $R(\pm\pi/2)$, with rotation operator $R_y$ used for odd numbers of qubits and $R_z$ used for even numbers.

\subsection{Synthesizing Multi-Qubit Gates with Multi-Mode Driving}
To generate programmable $n$-body interactions using global multi-mode drives, we will expand upon the template circuit introduced in Fig. \ref{fig:multiqubit}. We seek to overcome two limitations of existing approaches: (1) the multi-qubit unitaries $U_{MQ}$ in Fig. \ref{fig:multiqubit} represent fully-entangling  M\o lmer-S\o rensen interactions with uniform couplings. In theory, these can only approximately be achieved using global beams with single-mode drives due to coupling to spectator modes. (2) For an $N-$qubit system, the above approach cannot generate $n$-body interactions between a subset of qubits $n < N$ using global entangling beams.

Our approach for implementing $n$-body gates uses the arbitrary Ising interactions shown in Sec. \ref{sec:arbising} to enact the unitary operations $U_{MQ}$ in Fig. \ref{fig:multiqubit}. Within this framework, limitation (1) above may be circumvented by using multi-mode global drives to generate a pure all-to-all Ising interaction. This corresponds to choosing the multi-mode drive frequencies and amplitudes such that $c_k=1$ for the center-of-mass mode in Eq. \ref{eq:LinearCombinationJks}, and $c_k=0$ otherwise. Similarly, limitation (2) may be addressed by tailoring the Ising couplings so that only the desired subset of qubits $n < N$ participate in the $n$-body interaction. This corresponds to setting $J_{ij}$ constant for couplings between the $n$ participating qubits, and setting $J_{ij}=0$ otherwise. Since arbitrary Ising interactions may be realized following the techniques in Sec. \ref{sec:arbising}, any $n$-body gate implementation is always accessible using this framework.

\begin{figure*}[hbtp]
\centering
\includegraphics[width=.6\textwidth]{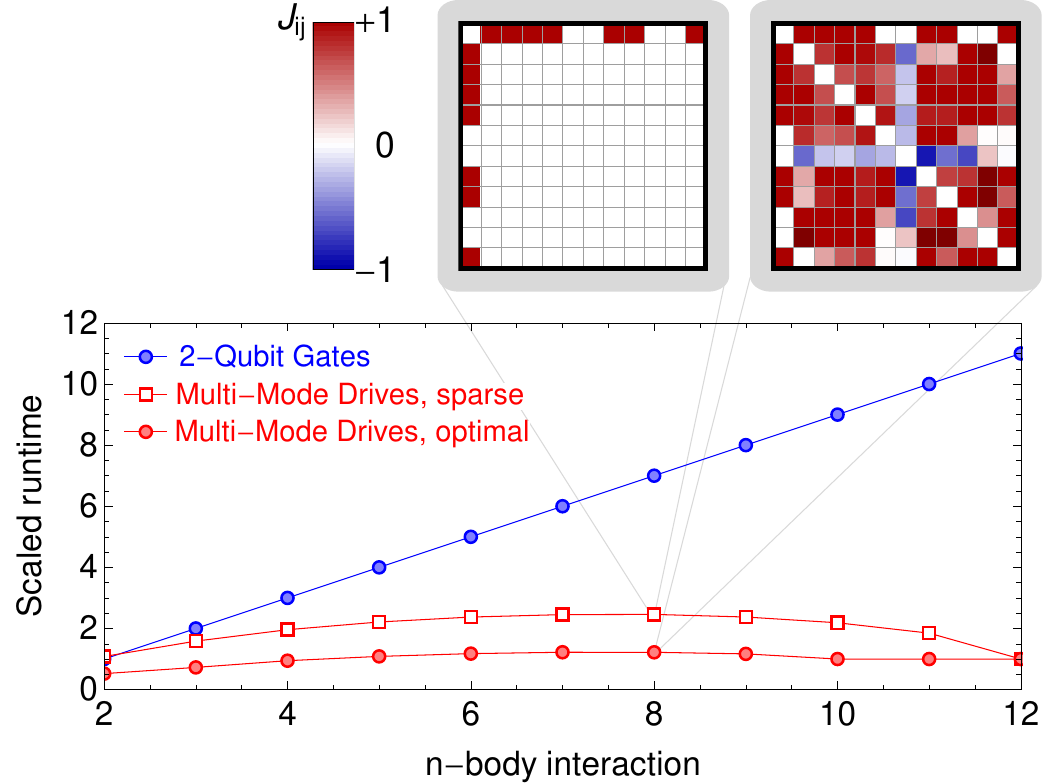}
\caption{
Runtimes for implementing $n$-body interactions in a chain of 12 ions. Each point represents the average runtime when $n$ out of 12 ions participate. Sequential two-qubit gates (blue circles) require linearly increasing runtimes. Shorter runtimes are achievable using multi-mode driving (red), either by engineering the same interaction matrix as in the two-qubit approach (open squares), or by implementing the $J_{ij}$ matrix that minimizes the runtime (filled circles). Example matrices for each case are shown as insets.
}
\label{fig:multiQubitData}
\end{figure*}

In Fig. \ref{fig:multiQubitData}, we compare the runtimes for implementing $n-$body interactions with up to $N=12$ ions. In the direct two-qubit gate approach, the unitary operators $U_{MQ}$ may be decomposed as sets of $n-1$ controlled-NOT gates between the top qubit and all other participating qubits \cite{nielsen2010quantum}. As a result, the direct two-qubit gate approach exhibits a linear scaling of runtime with $n-$body system size. In contrast, when multi-mode global drives are used to generate these same $U_{MQ}$ operators, we observe much more favorable runtimes which are approximately constant for increasing system size. This scaling is exactly constant for the $n=N$ case, since the $U_{MQ}$ operators may be implemented by driving a center-of-mass interaction for a single timestep, independent of system size.

Due to the structure of the example circuit in Fig. \ref{fig:multiqubit}, only the Ising interactions involving the top qubit matter; all others are canceled by the $U_{MQ}$ and subsequent $U^\dagger_{MQ}$ operations. For this reason, the left inset of Fig. \ref{fig:multiQubitData} has zeroed all Ising interactions unless they involve the top qubit and another participating qubit. However, this Ising interaction does not lead to the fastest runtime in a multi-mode drive implementation. By keeping the $J_{1j}$ and $J_{i1}$ interactions as needed for the $n$-body gate, we solve for the remaining set of Ising couplings that provides the minimum possible runtime. As shown in Fig. \ref{fig:multiQubitData}, this results in further improvement in the scaled runtime (filled red circles) compared with naive multi-mode drives (open squares) and direct two-qubit gates (blue circles).


\section{Quantum Fourier Transform}
\label{QFT}

\begin{figure*}[t]
\centering
\begin{quantikz}[row sep=0.2cm,column sep=0.3cm]
& \gate[4]{QFT} &\\
& \ghost{Z}               &\\
& \ghost{Z}              &\\
& \ghost{Z}               &
\end{quantikz}
=
\begin{quantikz}[row sep=0.2cm,column sep=0.3cm]
&\gate{H}  & \ctrl{3}   &          &            &          &            &          &\\
&\ghost{Z} & \gate{R_2} & \gate{H} & \ctrl{2}   &          &            &          &\\
&\ghost{Z} & \gate{R_3} &          & \gate{R_2} & \gate{H} & \ctrl{1}   &          &\\
&\ghost{Z} & \gate{R_4} &          & \gate{R_3} &          & \gate{R_2} & \gate{H} &
\end{quantikz}
\caption{
Traditional quantum circuit for implementing the Quantum Fourier Transform on a 4-qubit system. The $\mathcal{O}(N^2)$ $R_k$ gates denote controlled phase rotations by angle $2\pi/2^k$, and $H$ is the Hadamard gate. Swap gates at the end of the circuit are omitted for clarity.
}
\label{fig:qft}
\end{figure*}
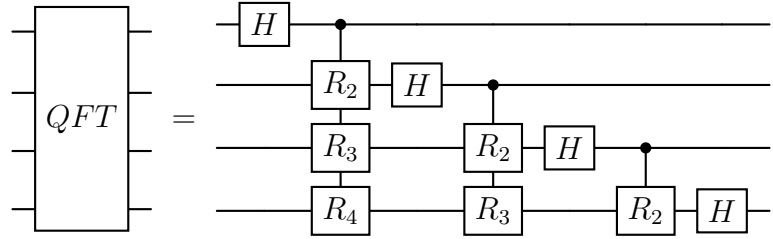

The Quantum Fourier Transform (QFT) plays a central role in key quantum algorithms, such as Quantum Phase Estimation and Shor's algorithm \cite{nielsen2010quantum}. A traditional QFT circuit, shown in Fig. \ref{fig:qft}, is constructed from single-qubit Hadamard operations and $\mathcal{O}(N^2)$ controlled phase rotations applied to an $N-$qubit system. In this section, we will show how multi-mode global driving allows for a quadratic reduction in the circuit depth, reducing the number of entangling operations from $\mathcal{O}(N^2)$ to $\mathcal{O}(N)$. In addition, we will show that the scaled runtimes for implementing the full and approximate QFTs are also reduced when using multi-mode drives, compared to standard two-qubit gate approaches.

\subsection{Comparison to prior work}
Several prior works have proposed techniques for synthesizing the QFT using global entangling operations. In \cite{martin2020digital}, for instance, the authors build upon Digital-Analog Quantum Computation \cite{parra2020digital} to decompose the controlled-phase rotations into analog evolution blocks interspersed with pairs of local spin flips. Just as in Ref. \cite{parra2020digital}, this approach requires $\mathcal{O}(N^2)$ analog blocks for each set of controlled rotations appearing in the QFT circuit. Similarly, Ref. \cite{bassler2023synthesis} expresses the QFT using multi-qubit gates, along with a linear program to determine the time-optimal sequence of single-qubit flips between analog evolution blocks. 

Further work \cite{maslov2018use} shows how the QFT may be compiled using global operations and single-qubit rotations, optimized for the interactions native to trapped-ion hardware platforms \cite{maslov2017basic}. Under this approach, the required interaction strength between qubits should fall exponentially with distance, as $J_{ij} \sim 2^{-|i-j|}$. However, since power-law interactions $J_{ij} \sim 1/|i-j|^\alpha$ are much more natural for trapped-ion systems \cite{monroe2021programmable}, refs. \cite{maslov2018use} and \cite{nam2015structural} focus on such approximate implementations of the QFT rather than on exact synthesis. For specific ranges of power-law decay exponent $\alpha$, these studies find that QFT implementation fidelities in excess of 90\% may be recovered.

\subsection{Implementing the QFT with Multi-Mode Global Drives}
\label{sec:qftimplementation}
In our approach to implementing the QFT operation, we combine the framework introduced by Ref. \cite{maslov2018use} with the arbitrary Ising interactions described in Sec. \ref{sec:arbising}. First, we decompose the QFT in terms of single-qubit rotations and global entangling operations. Then, we use multi-mode global driving to directly create the required $J_{ij}$ couplings for implementing the QFT. In particular, since multi-mode drives can generate arbitrary patterns of couplings, they may be tailored to exactly produce the exponential decay $J_{ij} \sim 2^{-|i-j|}$ that was inaccessible to earlier approaches with global drives. 

\begin{figure*}[t]
(a)

\begin{center}
\begin{quantikz}[row sep=0.2cm,column sep=0.3cm]
& \ctrl{1}   & \ghost{R_x(-\frac{\pi}{2^k})}\\
& \gate{R_k} &
\end{quantikz}
=
\begin{quantikz}[row sep=0.2cm,column sep=0.3cm]
&\gate{Y(-\frac{\pi}{2})} & \gate[2]{XX(\frac{\pi}{2^{k+1}})} & \gate{X(-\frac{\pi}{2^k})} & \gate{Y(\frac{\pi}{2})} &\\
&\gate{H}                 &                                   & \gate{X(\frac{\pi}{2^k})}  & \gate{H}                &
\end{quantikz}
\end{center}
(b)

\begin{center}
\begin{adjustbox}{width=\textwidth}
\begin{quantikz}[row sep=0.2cm,column sep=0.3cm]
& \gate[4]{QFT} &\\
& \ghost{Y(\frac{\pi}{2})}               &\\
& \ghost{Y(\frac{\pi}{2})}              &\\
& \ghost{Y(\frac{\pi}{2})}               &
\end{quantikz}
=
\begin{quantikz}[row sep=0.2cm,column sep=0.3cm]
&\gate{H} & \gate{Y(-\frac{\pi}{2})} &\gate[4]{U_{QFT}} & \gate{X(-\frac{7\pi}{16})} & \gate{Y(\frac{\pi}{2})}  &                   &                                         &                          &                   & & &\\
&\gate{H} &                          &                  & \gate{X(\frac{\pi}{4})}    & \gate{Y(-\frac{\pi}{2})} & \gate[3]{U_{QFT}} &       \gate{X(-\frac{3\pi}{8})} & \gate{Y(\frac{\pi}{2})}  &                   & & &\\
&\gate{H} &                          &                  & \gate{X(\frac{\pi}{8})}    &                          &                   & \gate{X(\frac{\pi}{4})}   & \gate{Y(-\frac{\pi}{2})} & \gate[2]{U_{QFT}} & \gate{X(-\frac{\pi}{4})} & \gate{Y(\frac{\pi}{2})} &\\
&\gate{H} &                          &                  & \gate{X(\frac{\pi}{16})}   &                          &                   & \gate{X(\frac{\pi}{8})}   &                          &                   & \gate{X(\frac{\pi}{4})} & &
\end{quantikz}
\end{adjustbox}
\end{center}
\caption{
(a) Controlled $R_k$ gates may be decomposed into single-qubit rotations around the $X$ and $Y$ axes, and a single $XX$ interaction that depends on the desired $k$. (b) Applying the decomposition in (a) to the traditional QFT circuit in Fig. \ref{fig:qft} generates an equivalent QFT operation using only $\mathcal{O}(N)$ global unitaries $U_{QFT}$ (see text for definition).
}
\label{fig:qftdecomp}
\end{figure*}

Our full decomposition of the QFT operator is shown in Fig. \ref{fig:qftdecomp}. We begin by isolating a single controlled-phase rotation $R_k$ (Fig. \ref{fig:qftdecomp}a). This gate is decomposed into single-qubit rotations and an $XX$ interaction, which corresponds to evolution under the Ising Hamiltonian
\begin{equation}
    XX(\theta)=e^{-i\theta \sigma_1^x\sigma_2^x}=e^{-i J_{12} \sigma_1^x\sigma_2^x t}
\end{equation}
where we may define the rotation angle $\theta = J_{12}t$. Thus, each of the controlled phase rotations in the QFT operator between ions $i$ and $j$ (Fig. \ref{fig:qft}) may be individually mapped to single-qubit rotations and Ising interactions of the form $J_{ij}$.

In Fig. \ref{fig:qftdecomp}b, we demonstrate the full decomposition of the QFT operation into single-qubit rotations and $\mathcal{O}(N)$ global unitary operators $U_{QFT}$. Comparing the structure of the controlled rotations $R_k$ in Figs. \ref{fig:qft} and \ref{fig:qftdecomp}a, we define the $U_{QFT}$ operator as an Ising-type analog evolution:
\begin{equation}
    U_{QFT}=\exp\left[-i\sum_{i<j} J_{ij} \sigma_i^x \sigma_j^x t\right]
\label{eq:uqft}
\end{equation}
where $J_{ij}t=\pi/4\cdot2^{-|i-j|}$ when $i$ or $j$ is a control qubit, and $J_{ij}=0$ otherwise. For instance, the three $U_{QFT}$ operators shown in Fig. \ref{fig:qftdecomp}b should be evolved with Ising couplings
\begin{equation}
\nonumber
    (J_{ij}t)_1 = \begin{pmatrix}
        0 & \frac{\pi}{8} & \frac{\pi}{16} & \frac{\pi}{32}\\
        \frac{\pi}{8} & 0& 0& 0 \\
        \frac{\pi}{16} & 0& 0& 0 \\
        \frac{\pi}{32} & 0& 0& 0
    \end{pmatrix}~~~;~~~
    (J_{ij}t)_2 = \begin{pmatrix}
        0 & 0& 0& 0 \\
        0 & 0 &\frac{\pi}{8} & \frac{\pi}{16}\\
        0  & \frac{\pi}{8} & 0 & 0\\
        0 & \frac{\pi}{16} & 0& 0
    \end{pmatrix}~~~;~~~
     (J_{ij}t)_3 = \begin{pmatrix}
        0 & 0& 0& 0 \\
        0 & 0& 0& 0 \\
        0  & 0 & 0 & \frac{\pi}{8}\\
        0 & 0 & \frac{\pi}{8}& 0
    \end{pmatrix},
\end{equation}
which are all achievable following the arbitrary Ising protocols introduced in Sec. \ref{sec:arbising}.

\begin{figure*}[t]
\centering
\includegraphics[width=.6\textwidth]{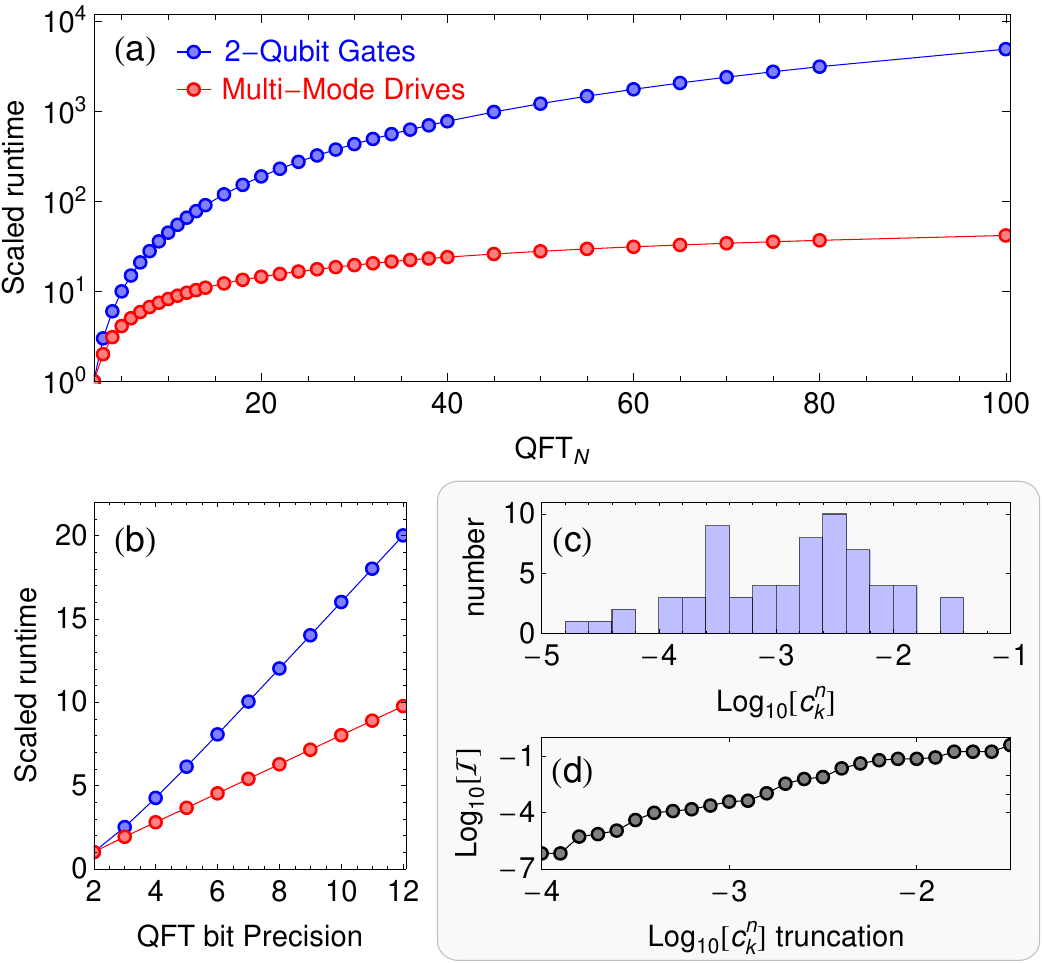}
\caption{
(a) Runtimes for implementing the Quantum Fourier Transform on $N$ ions using sequential two-qubit gates (blue) versus multi-mode driving (red). (b) Runtimes of the approximate QFT applied to a 12-ion system. Shorter runtimes are achievable for both direct two-qubit gates and multi-mod drives when lower precision is required. (c) When implementing multi-mode drives, many coefficients $c_k^n$ become small compared to typical experimental precision. (d) Truncating the $c_k^n$ coefficients below a certain threshold leads to small infidelities $\mathcal{I}$ when implementing the QFT.
}
\label{fig:qftscaling}
\end{figure*}

We quantify the performance of this QFT decomposition in Fig. \ref{fig:qftscaling}, by comparing the scaled runtimes of this approach and the traditional QFT. For a standard QFT implementation (Fig. \ref{fig:qft}), each of the $N$ qubits serves as a control for a series of $R_k$ gates, whose combined runtime scales as $\mathcal{O}(N^2)$. Hence, the overall runtime for a standard QFT circuit grows quadratically with system size, shown as the blue points in Fig. \ref{fig:qftscaling}a. For multi-mode global drives (red points in Fig. \ref{fig:qftscaling}a), we find polynomially faster runtimes which scale as the square root of system size, $\mathcal{O}(N^{1/2})$. In this approach, the runtime cost of implementing the arbitrary Ising matrices needed for the $U_{QFT}$ operators (Eq. \ref{eq:uqft}) decreases with system size as $\sim 1/\sqrt{N}$, and $N$ implementations of $U_{QFT}$ are required for the full QFT circuit.

We also consider implementations of the \emph{approximate} QFT \cite{nielsen2010quantum}, which reduces the number of entangling gates by keeping only the controlled phase rotations larger than a threshold angle. For $b-$bit precision, this needs only $\mathcal{O}(bN)$ controlled rotation gates rather than $\mathcal{O}(N^2)$ for the standard QFT. To implement the approximate QFT to $b-$bit precision using multi-mode global drives, the $J_{ij}$ couplings used for each iteration of the $U_{QFT}$ operator (Eq. \ref{eq:uqft}) should be kept only when $|i-j| < b$. Under these conditions, we compare the runtime performance of using direct two-qubit gates (blue) and multi-mode drives (red) in Fig. \ref{fig:qftscaling}b. We observe linear scaling in both cases, with a notably smaller slope for the multi-mode drive approach.

Finally, we consider an alternative approximation for the QFT operation, where we truncate the $c_k^n$ coefficients comprising the multi-mode drive if they fall below a threshold value. In Fig. \ref{fig:qftscaling}c, we show a histogram of the $c_k^n$ coefficients required for implementing the first $U_{QFT}$ operator during a full 12-ion QFT. We observe that nearly half of these coefficients are at the level of $0.001$ or below and contribute minimally to the unitary dynamics. Indeed, as demonstrated in Fig. \ref{fig:qftscaling}d, QFT implementation fidelities in excess of $99.99\%$ remain accessible even when half of the $c_k^n$ coefficients are eliminated. Such observations are key when considering practical implementations of multi-mode drives using imperfect experimental hardware.

\section{Conclusions}
\label{conclusion}
In this work, we have introduced a universal framework for synthesizing trapped-ion quantum logic operations using global entangling beams and local qubit rotations. This framework leverages the ease with which multiple independent frequency components may be added to global laser beams, allowing for tailored couplings to the available vibrational modes of the trapped-ion lattice. Although such global drives can only control up to $(N-1)$ of the $N(N-1)/2$ possible qubit-qubit couplings, interspersing these drives with local rotations enables fully universal control over the resulting Hamiltonian dynamics. As example applications, we have demonstrated the synthesis of Ising-type Hamiltonians with arbitrary spin-spin couplings, multi-body interactions beyond standard two-qubit gates, and an implementation of the Quantum Fourier Transform, all using global beams and local rotations.

Moreover, this multi-mode global drive framework provides more degrees of freedom than the number of qubit-qubit couplings. This allows for a set of drive parameters to be chosen that minimizes the overall runtime of the decomposed quantum circuit. We showed how this problem may be recast in terms of a linear program so that standard algorithms could quickly find an optimal solution. Our analysis of multi-mode drives has demonstrated that significant speedups are possible using this approach compared to direct implementations using two-qubit gates, and that quadratic speedups are achievable in many cases. In fact, for all examined applications, we have not found an instance where the direct two-qubit gate approach could be executed with shorter runtime than an optimized multi-mode global drive.

Experimental implementation of this multi-mode global drive protocol will require several specific considerations. For instance, generating the mode-coupling weights $c_k^n$ may practically be accomplished by passing a global beam through an acousto-optic modulator (AOM) driven with multiple independent rf tones. As shown in \cite{shapira2020theory,shapira2023fast,shapira2023programmable}, the number of required tones scales linearly with ion number $N$, while the overall Rabi frequency split amongst the tones scales as $\sqrt{N}$. Drifts in the amplitude of these tones, due to laser or AOM noise, may lead to an imperfect set of $c_k^n$ weights and would manifest itself as an infidelity in the applied $J_{ij}$ couplings. Likewise, unexpected motional-mode frequency drifts would also lead to a sub-optimal set of $c_k^n$ weights and result in lower fidelity operations. These observations suggest that significant attention should be given towards laser intensity stabilization \cite{wang2020reduction}, rf stabilization \cite{johnson2016active}, and frequent characterization of the motional modes. Further robustness to motional mode drift may be also achieved by applying additional frequency tones to satisfy additional constraints on the ions' phase space trajectories, following the approach in Ref. \cite{shapira2018robust}. The ultimate limitation of implementing multi-mode global drives will be the requirement to resolve all vibrational mode frequencies, which sets a minimum gate time $T > \pi/\Delta\omega_\text{min}$, with $\Delta\omega_\text{min}$ the minimum frequency splitting between modes. For a 30-ion chain of $^{171}$Yb$^{+}$ ions and standard trap parameters, we estimate a minimum gate time of $T > 300~\mu$s, which is reasonable compared to the $T \gtrsim 1$ ms gates common in adiabatic quantum simulation experiments \cite{monroe2021programmable}.

Within this framework, we expect that the efficiency of further classes of quantum circuits may be improved upon, beyond the Ising models, $n$-body gates, and QFT circuits explored here. For instance, future theoretical directions include the implementation of quantum error correction protocols (which rely upon multi-body operators applied to subsets of qubits), decomposition of algorithms such as quantum phase estimation and period-finding using multi-mode global drives, and efficient block encodings of Hermitian operators for Hamiltonian simulation. Compared to the traditional approaches using direct two-qubit gates, this work provides an alternate paradigm for synthesizing quantum logic operations of direct relevance, without the steep experimental overhead associated with individually-addressed entangling beams.


\section*{Acknowledgments}
The author thanks Antonis Kyprianidis for early discussions. This work was supported in part by the Gordon and Betty Moore Foundation, GBMF \#12963 and by the National Science Foundation under Grant No. PHY-2412878. The IU Quantum Science and Engineering Center is supported by the Office of the IU Bloomington Vice Provost for Research through its Emerging Areas of Research program.


\section*{References}
\bibliographystyle{unsrt}
\bibliography{refs}

\end{document}